\documentclass[ukenglish,notitlepage,a4paper,12pt]{article}
\usepackage{sw20elba}
\usepackage[dvips]{graphicx}

\begin{document}

\title{\textbf{Causal Implication and the Origin of Time Dilation}}
\author{\textbf{George Jaroszkiewicz} \\
School of Mathematical Sciences, University of Nottingham,\\
University Park, Nottingham NG7 2RD, UK}
\date{Sunday 30$^{th}$ July 2000}
\maketitle

\begin{abstract}
We discuss the emergence of time dilation as a normal feature expected of
any system where a central processor may have to wait one or more clock
cycles before concluding a local calculation. We show how the process of
causal implication in a typical Newtonian cellular automaton leads naturally
to Lorentz transformations and invariant causal structure.
\end{abstract}

\section{ Introduction}

In this paper we develop the idea that time dilation of the sort encountered
in special relativity occurs naturally in any universe which behaves like a
cellular automaton. This theme was discussed by Minsky nearly two decades
ago \cite{MINSKY-82}, but at that time the ``wrong'' dilation factors
emerged and the idea seems not to have been taken further. We obtain our
results by focussing on the frame of references employed, rather than on
wave packets as was the case of the Minsky paper, and this makes the
conclusions more generic.

Our discussion is based on an approach to causality introduced by us \cite
{JAROSZKIEWICZ-99}, in which time emerges as an index of causal implication
as carried out by a Theorist standing outside of space-time. By this we mean
the following. Between times $n$ and $n+1$ of the Theorists's internal
discrete clock time (referred to as \emph{physiotime} in \cite
{JAROSZKIEWICZ-99}), the Theorist works out the $n+1^{th}$ \emph{causal
implication}, which is the maximal set of future data values logically
inferable from all the information available to the Theorist at time $n$,
using the given laws of physics.\ This process is repeated, with $n$
replaced by $n+1$, \emph{ad infinitum}. This gives a natural dating process
to events.

From this point of view the Theorist behaves effectively as a Turing
computing machine, the input of which involves the dynamical data in some of
the events or cells of space-time, including conventional past and future
events. If at time $n\;$a given cell is blank (i.e. unwritten) and the local
rules of the cellular automaton allow it, then the Theorist can make a
causal implication at that cell, writes in the value determined by the
dynamics, and dates the cell $n+1$. Otherwise the cell is left alone until
some possibly future physiotime. This model reconciles the notion of a static block
universe with the process view of time, where the present is a real,
distinguishable moment.

Our approach utilises only the local discrete topology of cellular automata 
(the relationship between adjacent cells),
which makes it independent of the details of the dynamical rules relating
state values in adjacent cells. Therefore, the results are generic and
should find their analogues in a wide variety of cellular automata. Our
conclusion is that the notion of time dilation is not specific to Minkowski
space-time but occurs also as a generic feature of systems such as modern
computers where information is processed by a central processing unit
governed by an internal clock. On occasions, the processor may have to wait
various clock cycles before certain local computations can be completed,
simply because necessary information is not yet available, and this is
equivalent to time dilation.

\section{Discrete space-time}

Cellular automata are of interest to physicists for a number of reasons \cite
{WOLFRAM:86}. First, they deal with discrete space-time, which is currently
a fashionable notion. Second, they generate in a completely natural way
intrinsic structures analogous to the light-cone structure of special
relativity. Third, they can provide good approximations to the differential
equations of conventional physics. From our point of view, however, we prefer to 
put it the other way around and regard conventional physics as a good
approximation, in an appropriate continuum limit, to some discrete
space-time cellular automaton.

In conventional cellular automata, such as discussed by Minsky \cite
{MINSKY-82}, there is an external absolute time $n$ which regulates the
dynamical evolution in a strict way. At each click of this time, each cell
(or \emph{event} in the terminology of \cite{JAROSZKIEWICZ-99}) $C^m$
carries a dynamical variable (the \emph{state }of the event in the
terminology of \cite{JAROSZKIEWICZ-99}) which has value $\psi _n^m$ at that
time. For such automata, there will be some deterministic law or rule such
that the value $\psi _{n+1}^m$ of the state in $C^m$ at time $n+1$ is
completely determined (or \emph{causally resolved }in the terminology of 
\cite{JAROSZKIEWICZ-99}) by knowledge of some or all of the values $\left\{
\psi _n^m:-\infty <m<\infty \right\} $ at an earlier time $n.$

In our approach we have such a structure, but it is modified in an important
way as follows. First, there will in general also be rules (called links in
the terminology of \cite{JAROSZKIEWICZ-99}) which relate cell values at
different values of $n$ and $m$. Examples are 
\begin{equation}
\psi _{n+1}^m+\psi _n^{m+1}+\psi _n^m+\psi _n^{m-1}+\psi
_{n-1}^m=1,\;\;\;\;-\infty <m<\infty ,\;\;\;\psi _i^j\in C  \label{one}
\end{equation}
and 
\begin{equation}
g_{n+1}^mg_n^{m+1}g_n^mg_n^{m-1}g_{n-1}^m=e,\;\;\;\;\;\;g_i^j\in SU(2),
\label{two}
\end{equation}
where $e$ is the identity element of the group. These two quite different
links have the same local discrete topology, in the sense that the pattern
of $n^{\prime }s$ and $m^{\prime }s$ is the same. These links will generate
identical causal structures, even though they represent quite different
dynamical systems. In the following we shall discuss cellular automata with two indices, 
$m$ and $n$ as above, but our conclusions are general.

The difference between our approach and conventional cellular automata is
that we allow the external Theorist (the central processor) to work with
cells which may be scattered over the $m-n$ domain in a more general way
than just lined up at equal values of $n$, say, corresponding to
conventional ``planes of simultaneity''. We shall show how standard Lorentz
time dilation and Fitzgerald length contraction emerges using these ideas,
based on the local discrete topology implied by $\left( \ref{one}\right) $
and $\left( \ref{two}\right) .$ This topology was not chosen by accident, as
it emerges naturally in simple discretisations of the Klein-Gordon equation
as shown below.

\section{Causal implication}

Given the continuous time equation 
\begin{equation}
\partial _t^2\psi \left( t,x\right) -\partial _x^2\psi \left( t,x\right)
+\mu ^2\psi \left( t,x\right) =0,
\end{equation}
in $1+1$ space-time dimensions, with the speed of light and Planck's
constant chosen to be unity, a simple discretisation of space-time defined
by 
\begin{eqnarray}
t_n^m &\equiv &nT,\;\;\;x_n^m\equiv mL,  \nonumber \\
\psi _n^m &\equiv &\psi \left( t_n^m,x_n^m\right)
\end{eqnarray}
where $T$ and $L$ are scale constants satisfying $T/L=1,$ gives the link
equations 
\begin{equation}
\psi _{n+1}^m+\psi _{n-1}^m-\psi _n^{m+1}-\psi _n^{m-1}+\mu ^2T^2\psi
_n^m+O\left( T^4\right) =0.
\end{equation}
Neglecting the higher order terms, we arrive at an equation of the type 
\begin{equation}
\psi _{n+1}^m=\mathcal{F}\left( \psi _n^m,\psi _n^{m+1},\psi _n^{m-1},\psi
_{n-1}^m\right) ,  \label{three}
\end{equation}
which has the same local discrete topology as $\left( \ref{one}\right) $ and 
$\left( \ref{two}\right) .$

In our approach, such an equation can only be resolved (or worked out) if at
Clock Time (
$\equiv $ physiotime	   ) $p$, the Theorist ($%
\equiv $ central processor) actually knows what the state values $\psi
_n^m,\;\psi _n^{m+1},\psi _n^{m-1}$ and $\psi _{n-1}^m\;$in the cells $%
C_n^m,C_n^{m+1},C_n^{m-1},C_{n-1}^m$ are. If these are not available, the
Theorist cannot complete this calculation and must wait.

If these values had in fact been computed at previous Clock Times $a,b,c,d$
respectively then the value $\psi _{n-1}^m\;$on the left hand side of $%
\left( \ref{three}\right) $ could be evaluated or resolved, but no earlier
than at Clock Time $p=\max (a,b,c,d).\;$ In such a case the calculation is
completed and dated, and forms part of the $\left( p+1\right) ^{th}$
implication. This provides a temporal ordering or dating over space-time.

\subsection{The algorithm}

Specifically, the algorithm for causal implication in such a model is as
follows.

\begin{enumerate}
\item  At Clock Time $p$, the Theorist inspects every cell $C_i^j$ in
space-time and marks each one down for possible implication, or not, as
follows:

\begin{enumerate}
\item  if a particular cell $C_i^j$ already has a date, this means that the
cell value ($\equiv $ event state)\ $\psi _i^j$ has already been evaluated
at that cell at some previous Clock Time, and the cell is left alone and the
cell value never changes;

\item  if a particular cell $C_i^j$ has no date, then the Theorist must look
at cells $C_{i-2}^j$, $C_{i-1}^j$, $C_{i-1}^{j+1}$, $C_{i-1}^{j-1}.$ If one or more of
these cells has no date then $C_i^j$ cannot be resolved at Clock Time $p$. If
however, each of the cells $C_{i-2}^j,C_{i-1}^j,C_{i-1}^{j+1},C_{i-1}^{j-1}$ already has
a date (which necessarily must be less or equal to $p)$ then $C_i^j$ is
marked down for resolution:
\end{enumerate}

\item  The Theorist now resolves ($\equiv $ evaluates) each cell which has
been marked down for resolution in the first step of this algorithm and
dates it with time $p+1$.
\end{enumerate}

We note the following:
\begin{enumerate}
\item
If the $m-n$ plane is infinite the Theorist may take an infinite amount of
physiotime to inspect each cell. This is not regarded as a problem. In
practice, physicists only deal with finite regions of space-time, and we may
idealise this to the infinite extent situation.

\item
The actual duration of a Clock Time interval is not significant here. Ticks
of Clock Time only occur when a given process of causal implication has been
completed by the Theorist.

\item
To paraphrase the words of Omar Al-Khayyam, who was a mathematician as well
as a poet, once the moving finger of the Theorist has written in a cell, it
moves on and \emph{never }rewrites that cell.

\item
A cell which is being resolved cannot be used in the process of causal
resolution for any other cell during that particular implication process.
The role of cells is entirely \emph{classical }here, in that a given cell
can only play one role at any given Clock Time. Either it is being resolved,
does nothing, or else is being used in the resolution of other cells.

\item
Conventional spreadsheets are very convenient for analysing cellular
automata along the above lines .
\end{enumerate}

\textbf{Example:}
Suppose our initial data set $\sigma _{0\;}$consists of cells with given
values $\psi _{-1}^m$ and $\psi _0^m$ for $-\infty <m<\infty .\;$ Using $%
\left( \ref{three}\right) $ we can work out the \emph{first implication} $%
\sigma _1$, given by 
\begin{equation}
\sigma _1=\left\{ \psi _1^m:\;\;\psi _1^m=\mathcal{F}\left( \psi _0^m,\psi
_0^{m+1},\psi _0^{m-1},\psi _{-1}^m\right) ,\;-\infty <m<\infty \right\} .
\end{equation}
Here we are using events and event states interchangeably in our notation,
and it should be clear what is meant.

Once we have completed this task, we are in a position to repeat the process
of implication, but with the difference that the data set used for the \emph{%
second implication} $\sigma _{2\;}$now involves the state values $\psi _1^m\;
$and $\psi _0^m$ for $-\infty <m<\infty .$

Clearly this process can be continued indefinitely. There is no difference
in this example between our approach and that taken in conventional cellular
automata, but this is not true in general.

\section{Inertial frames}

The initial data set $\sigma _0$ in the above example would be naturally
identified as relating to the fundamental rest frame $\mathcal{F}_0$ of the
cellular automaton space-time lattice, in which the dynamical rules are
specified.  If we wish to simulate inertial frames moving with respect to $%
\mathcal{F}_0$ we need to start with different initial data sets. We are
free to choose a number of subsets of the lattice for this, but some choices
are more natural than others. We shall work with a reasonable choice as
follows. First, we choose a frame velocity $v\equiv r/s$, where $r$ and $s$
are integers and $|r|<s$. In this approach, therefore, all velocities are
rational fractions of the speed of light (chosen to be unity).

We define our initial data set $\sigma _0\left( v\right) $ as shown in $%
Fig.\;1.\;$ In this diagram $n$ runs left to right and $m$ runs upwards.
Cells belonging to $\sigma _0\left( v\right) $ are labelled by zeros and
shaded, and contain given starting state values. State values are \emph{not}
shown in such a diagram.

By the process of causal implication, we may fill in cells on either side of
the zig-zag initial data set. We shall only discuss implications running to
the right, as this represents the forwards direction of time in our model.

\begin{figure}[!t]
\begin{center}
\includegraphics[width=470.0pt,height=486.0pt]{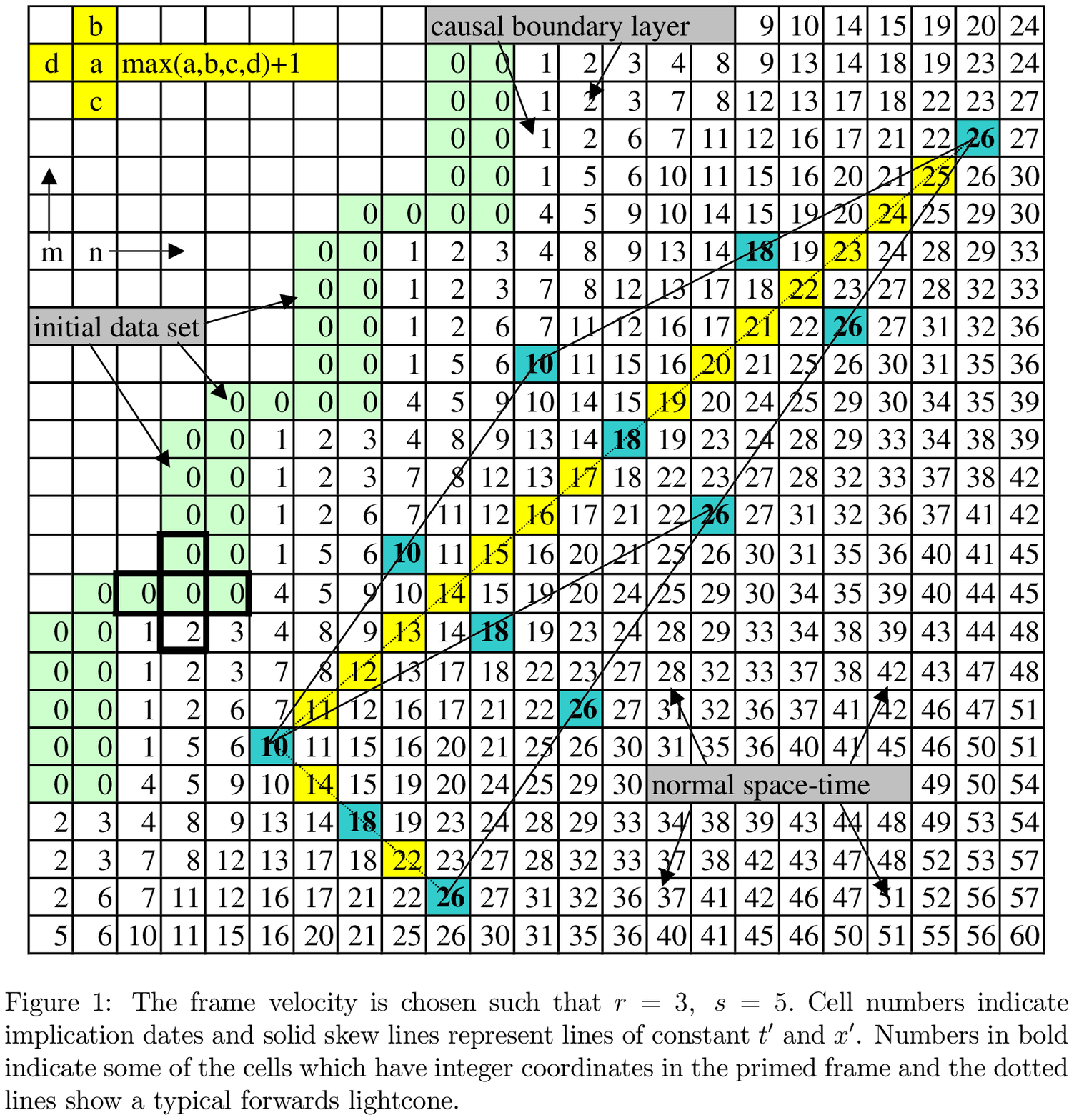}
\end{center}
\end{figure}

The pattern of implication settles down after several Clock cycles and
assumes a regularity in which a relativistic pattern may be discerned. In $%
Fig.\;1$ we show lines which connect cells which carry dates suggestive of
times and position coordinates in a frame $\mathcal{F}_5^3$ moving with
velocity $v=3/5$ relative to the rest frame $\mathcal{F}_0$ of the original
lattice.

These patterns are rather general for this particular cellular automaton and
for analogous initial data sets, so we now give a more general discussion,
which helps to illuminate features of $Fig.\,1$.

\section{The general case}

We may generalise the above example and consider transforming to a frame of
reference $\mathcal{F}_s^r$ moving with velocity $v\equiv r/s$ relative to $\mathcal{F}_0$, 
where $r$ and $s$ are integers
with the proviso $|v|<1$. Now the initial data set consists of a double
layer of zeros zig-zagging over the $n-m$ plane\ in a manner identical to
that shown in $Fig.\;1$, except now instead of five steps upwards followed
by three steps to the right, and so on, we have $s$ steps upwards followed
by $r$ steps to the right, and so on \emph{ad infinitum.\ }

With such an initial data set, experience soon confirms the following
pattern of implication, analogous to that shown in $Fig.\;1$. During the
initial phases of causal implication, there is something like a boundary
layer, starting with the first implication, during which the final patterns
begin to get established. Care should be taken to ensure that any discussion
concerning frames of reference does \emph{not} involve cells inside this
boundary layer. This is particularly the case if such a model is evaluated
using a spreadsheet. A spreadsheet is an excellent tool to use for
calculating such causality diagrams, as single changes to any cell or cells
in the initial data set are immediately translated throughout the rest of
the space-time, and the causal structure of light-cones can be very readily
discerned.

An important point concerns the inital data set. It is possible 
that such a set is a ``Garden of Eden" set \cite{WOLFRAM:86}, which means that it cannot be 
derived from the rules of the particular cellular automaton being discussed. In $Fig.\;1$
it will be observed that some of the inital data set cells appear to be inconsistent with the rules. The five cells shown with a bold border in $Fig.\;1.$ are an example where the right most cell cannot be an implication of the others, since the lowest one is dated two implcations later. However, a strict reading of the algorithm shows that the whole process is in fact consistent. 

We note that such a scenario occurs in the causal boundary layer. Once the process of implication has cleared this layer, no apparent clashes occur. This will be important in cases such as the discretised Klein-Gordon equations where the causal implication is reversible, i.e., causal implication can run in two or more directions depending on the boundary conditions.

The depth of the causal boundary layer will depend on the values of $r$ and $%
s$. It is remarkable that quite without any further input, the final pattern
beyond the boundary layer emerges as a result of some sort of
self-organisation within the boundary layer. What is clear from our results
is that complex behaviour can emerge from simple dynamical rules, but may
depend critically on starting conditions.

Beyond the causal boundary layer, the flow of time as indicated by the
implication dates assumes a regular pattern of behaviour, much like Minsky's
wave packet ``machines'' \cite{MINSKY-82}. Again, the periods discernible in
this pattern depend on $r$ and $s$.\ One immediate observation is that the
date of an implication in a cell does \emph{not} in general increase
strictly uniformly with ``coordinate time'' $n$, but over and above any
regular deviation, proceeds proportionately with $n\;$ and $m$.\ From now on
we shall assume we are beyond the causal boundary layer and that normal
space-time has stabilised.

By inspection of a number of examples, we are led to the following. First,
denote points in the \thinspace $m-n\;$plane by vectors 
\begin{equation}
\mathbf{x}_n^m \equiv \left( n,m\right) .
\end{equation}
Then
\begin{equation}
\mathbf{x}_p^0 \equiv \left( p,0 \right) \;\;\; p \;\;\mathrm{ an\ integer}
\end{equation}
represents a time-like vector whereas
\begin{equation}
\mathbf{x}_0^q\equiv (0,q),\;\;\;q \;\;   \mathrm{ an\ integer}
\end{equation}
represents a space-like vector.

Let 
\begin{equation}
\mathbf{x}_n^m\equiv \left( t_n^m,x_n^m\right) .
\end{equation}
Then 
\begin{equation}
t_n^m=n,\;\;\;x_n^m=m
\end{equation}
are time and space coordinates in our absolute rest frame $\mathcal{F}_0$,
in which the rules of the cellular automaton are defined.

The objective now is to transform to some new frame $\mathcal{F}_s^r$ with
coordinates 
\begin{equation}
t_n^{m\prime },\;\;x_n^{m\prime },
\end{equation}
where now 
\begin{equation}
t_m^{n\prime }=date(\mathbf{x}_n^m),
\end{equation}
where $date$ returns the implication date based on the given initial data
set $\sigma \left( r,s\right) $. The interpretation of $x_n^{m\prime }$
remains to be determined.

By inspection of a ``light-line'', representing the maximal flow of
causality along the diagonal $m=n$ in $\mathcal{F}_0$, we found in all
examples studied that the implication date $t_n^{m\prime }\;$ along such a
line satisfies the rule 
\begin{equation}
t_n^{m\prime }=t_n^m,  \label{light}
\end{equation}
assuming the zero of counting has been reset to occur inside normal
space-time. This simply amounts to subtracting off a suitable positive
constant from all dates in space-time.

However, since we also expect something like a Lorentz transformation to
take us from $\mathcal{F}_0$ coordinates to $\mathcal{F}_s^r$ we write 
\begin{eqnarray}
t_n^{m\prime } &=&\Gamma _s^r\left( t_n^m-\frac rsx_n^m\right) ,  \nonumber
\\
x_n^{m\prime } &=&\Gamma _s^r\left( x_n^m-\frac rst_n^m\right) ,  \label{L1}
\end{eqnarray}
where $\Gamma _s^r$ is some scale factor to be determined.

Along the above light-line, we may use $\left( \ref{light}\right) $ and the
relation 
\begin{equation}
t_n^m=x_n^m
\end{equation}
to find 
\begin{equation}
\Gamma _s^r=\frac s{s-r} = \frac 1{1-v},
\end{equation}
which is always positive.

A problem now emerges. All cells in $\mathcal{F}_0$ are labelled by integers 
$n$ and $m$. In fact, it is true the other way around. Given \emph{any }pair
of integers\ $\left( a,b\right) $ then there exists a unique cell $C_a^b$ in 
$\mathcal{F}_0\;$corresponding to this pair. However, this is \emph{not}
true of the frame $\mathcal{F}_s^r$. The coordinates $t_n^{m\prime }$ and $%
x_n^{m\prime }$ given by the rule $\left( \ref{L1}\right) $ are not in
general integers for arbitrary choice of integers $m$ and $n$. The reason
for this can be understood in more than one way. First, the transformation $%
\left( \ref{L1}\right) $ is a pseudo-rotation of a cubic lattice, and it is
obvious that such a transformation will not in general rotate one cubic
lattice exactly into another cubic lattice. This problem is one frequently
encountered in lattice gauge theories, and is addressed by an appeal to the
continuum limit, i.e., it is argued that loss of rotational invariance is
recovered in the limit of zero lattice spacing.

This argument does not help us here, and indeed, we would not wish to use
it. We are not interested in the continuum limit at this stage, and it is
essential for us to retain discreteness. Therefore, we have to get around
this problem in a different way.

A second way of looking at this issue is to recognise that in such a model,
only the fundamental rest frame $\mathcal{F}_{0\;}$ ``exists'' in an
meaningful way on the microscopic level, where the rules of the cellular
automaton are defined. Other frames of reference such as $\mathcal{F}_s^r$
are convenient fictions, which play a significant role on much larger
scales, when the continuum versions of relativity begin to hold.

In fact, only some cells in the fictitious lattice $\mathcal{F}_s^r$
transform under $\left( \ref{L1}\right) $ from some cells in $\mathcal{F}_0$%
. Other cells in $\mathcal{F}_s^r$ are convenient interpolations and need
not have any physical significance.

To determine which cells should be used in the transformation, consider the
vectors in $\mathcal{F}_0$ given by

\begin{equation}
\mathbf{m}\equiv \left( r,s\right) ,\;\;\;\;\mathbf{n}\equiv \left(
s,r\right) .
\end{equation}
Then consider those cells in $\mathcal{F}_0$ at the sites 
\begin{equation}
\mathbf{x}\equiv \left( t,x\right) =p\mathbf{n}+q\mathbf{m,}
\end{equation}
where $p$ and $q$ are integers. Then the coordinates $t\;$and $x$ of these
cells are integers given by 
\begin{eqnarray}
t &=&ps+qr  \nonumber \\
x &=&pr+qs.
\end{eqnarray}
This defines a sub-lattice of $\mathcal{F}_0$. With the transformation rule $%
\left( \ref{L1}\right) $ and dropping the sub- and super-scripts, we find 
\begin{equation}
\mathbf{x}^{\prime } \equiv \left( t^{\prime },x^{\prime }\right)
=\left( s+r\right) \left( p,q\right) ,
\end{equation}
i.e. 
\begin{eqnarray}
t^{\prime } &=&p(s+r)  \nonumber \\
x^{\prime } &=&q(s+r),  \label{int}
\end{eqnarray}
which are integers. Essentially, we have a transformation between a
sub-lattice of $\mathcal{F}_0\;$to a sub-lattice of $\mathcal{F}_s^r$.

The inverse transformation exists and is given by 
\begin{eqnarray}
t &=&\Gamma _{-s}^r\left( t^{\prime }+\frac rsx^{\prime }\right)  \nonumber
\\
x &=&\Gamma _{-s}^r\left( x^{\prime }+\frac rst^{\prime }\right) .
\end{eqnarray}
and we find that provided $t^{\prime }$ and $x^{\prime }$ satisfy $\left( 
\ref{int}\right) $ then we recover integer valued $t$ and $x.$

The light-cone structure of the cellular automaton is invariant to these
transformations. We find 
\begin{eqnarray}
t^{\prime 2}-x^{\prime 2} &=&\left( \frac{s+r}{s-r}\right) \left(
t^2-x^2\right)  \nonumber \\
&=&\left( \frac{1+v}{1-v}\right) \left( t^2-x^2\right) ,  \label{caus}
\end{eqnarray}
so that space-like, time-like, and light-like intervals in $\mathcal{F}_0$
appear as space-like, time-like and light-like intervals respectively in $%
\mathcal{F}_s^r.\;$ Causality therefore is invariant to these
transformations.

\section{Interpretation and discussion}

We note that we started off without any concept of Minkowski metric. Indeed,
the discrete topology of our cellular automaton rules is similar to those
found in cellular automata such as Conway's ``Game of Life'', which is quite
Newtonian in its operation. The essential difference is the role of causal
implication and the occasional need for the Central processor (the Theorist)
to delay causal resolution in a cell until information becomes available.

It is remarkable, therefore, that a special relativistic structure emerges
from such a Newtonian basis. In this approach, there \emph{is} an absolute
time underlying the running of the universe, but it is the Clock Time of the
Theorist, and not that of the space-time diagram per se. Proper time and
relativistic time dilation can be now understood as signals of this
relationship. It is also very clear from the model why time \emph{dilation}
occurs, and not time\emph{\ contraction. }Computers can always lose time,
but cannot go faster than their natural clock rate.\emph{\ }

Several points need to be addressed. First, what about Minsky's apparently
wrong time dilation factor? We note that the dilation factor $\Gamma _s^r$
that we found is not the conventional factor 
\begin{equation}
\gamma \left( v\right) \equiv \frac 1{\sqrt{1-v^2}}
\end{equation}
either. However, looking at the transformation equations $\left( \ref{L1}%
\right) $ and the causality invariance property $\left( \ref{caus}\right) $,
we see that the problem is one merely of an overall choice of scale of
units. It is only by choosing to apply the principle of special relativity,
i.e., that all inertial frames of reference should be formally equivalent,
that we would be led to rescale our definitions of $t^{\prime }$ and $%
x^{\prime }$ by the rule 
\begin{eqnarray}
t^{\prime } &\rightarrow &\tilde{t}\equiv \sqrt{\frac{1-v}{1+v}}t^{\prime } 
\nonumber \\
x^{\prime } &\rightarrow &\tilde{x}\equiv \sqrt{\frac{1-v}{1+v}}x^{\prime }.
\end{eqnarray}
This is something invoked for conventional macroscopic scale physics, but
on a microscopic level, there is no reason to do this. Therefore, we suggest
that Minsky was entirely correct in his original message.

Another problem is the fact that the transformations relate integer labelled
cells only for sub-lattices of $\mathcal{F}_0$ and $\mathcal{F}_s^r.$ This
is related to the need also to consider only rational fractions of the speed
of light. There are two answers which come to mind. Either we allow inertial frames 
moving or spatially rotated with respect to the fundamental frame to be distorted cubic lattices, or we appeal to physical scales. We may imagine that the
scales $T$ and $L$ discussed earlier relate to something like Planck scales,
i.e., we may suppose $T$ is of the order $10^{-44}$ seconds\ and $L$ is of
the order $10^{-35}$ metres. This would mean that there is considerable scope for
taking relatively large multiples of $T$ or $L$ (such as $s,r\sim 10^{10})$
and still leave such vast number of points per second or metre in the
lattice $\mathcal{F}_s^r$ for a continuum to provide a good approximation.

There may be regimes accessible to current physics which might permit
investigation of this last point, but we will leave investigation of this to
a future paper. In other words, there may be empirical consequences to these
ideas.

Four more points need to be discussed.

\begin{enumerate}
\item  In the model chosen and with the initial data set used in $Fig.\;1$,
causal implication would give resolution moving to the left as well as to
the right. This is because the dynamics is reversible. This is not a
problem, as such behaviour occurs in conventional mechanics. The Theorist
needs to decide which is the correct physical direction of time flow and
ignore other disconnected pieces of the causal implication.

\item  There is a fundamental frame, $\mathcal{F}_0$ in this model, in which
the lattice $\left( n,m\right) $ makes sense. Lest this be regarded as
contrary to the principles of relativity, two things should be considered: $%
(i)$ the Cosmic Background Radiation Field does provide a physically sound
marker for identifying a special local inertial frame at each point in the
Universe, and $\left( ii\right) $ if indeed relativity emerges as a
continuum approximation to a discrete theory such as the above, then
relativistic principles cannot be used to criticise the discrete theory.

\item  The use of a Theorist standing outside space-time may appear ad-hoc 
and without explanation, but if the universe is indeed a vast cellular
automaton, then it is quite feasible for complex structures to emerge within
it which behave for all intents and purposes as a Theorist. This by no means
invalidates Penrose's notion that non-computational physics must be involved
somewhere in the origin of consciousness \cite{PENROSE:94}. Our model
utilises only the end result, which is to suppose the existence of a
Theorist who behaves effectively like a Turing machine. This is a curious
reversal of the usual idea which advocates of artificial intelligence might
discuss, namely the question whether a Turing machine might ever behave to
all intents and purposes in the fashion of a human.

\item  The role of quantum mechanics in such a scheme remains to be
investigated and we will report on this presently.
\end{enumerate}

\section{Acknowledgments}

The author is indebted to Professor M. Minsky for communicating his ideas
and to Keith Norton for many interesting discussions on discrete time. This
paper is dedicated to the memory of Zo\"{e} Norton.


\newpage

\begin{thebibliography}{9}
\bibitem{MINSKY-82}  Marvin Minsky, \emph{Cellular Vacuum}, Int. J. Theor.
Phys., 21(6/8), pp537-551 (1982)

\bibitem{JAROSZKIEWICZ-99}  George Jaroszkiewicz, \emph{Discrete Spacetime:
Classical Causality, Prediction, Retrodiction and the Mathematical Arrow of
Time}, in \emph{First International Interdisciplinary Workshop: Studies on
the Structure of Time: from Physics to Psycho(Patho)Logy, }edited by V.
DiGesu, R. Buccheri and M.Saniga, 23-24 November 1999, CNR-Area della
Ricerca di Palermo, Sicily, published by Kluwer, Dordrecht. Also available
at gr-qc/0004026.

\bibitem{WOLFRAM:86}  Stephen Wolfram, \emph{Theory and Applications of
Cellular Automata}, World Scientific, 1986

\bibitem{PENROSE:94}  Roger Penrose, \emph{Shadows of the Mind}, Oxford
University press, 1994
\end{thebibliography}
\end{document}